\documentclass[journal]{IEEEtran}

\usepackage{times}
\usepackage{graphicx}
\usepackage{float}
\usepackage{lipsum}
\usepackage{balance}

\ifCLASSINFOpdf

\else

\fi

\begin{document}

\title{A Novel Architecture of Area Efficient FFT Algorithm for FPGA 
Implementation}

\author{\IEEEauthorblockN{Atin Mukherjee, Amitabha 
Sinha and Debesh Choudhury}\\ 
\IEEEauthorblockA{Neotia Institute of Technology, Management and Science 
\\ Department of Electronics and Communication Engineering \\ Diamond 
Harbour Road, Jhinga, PO - Amira, Sarisha, South 24 Parganas\\ Pin 
743368, West Bengal, India} 
}

\maketitle
\pagestyle{empty}
\thispagestyle{empty}

 \begin{abstract}
 Fast Fourier transform (FFT) of large number of samples requires huge 
hardware resources of field programmable gate arrays (FPGA), which needs 
more area and power. In this paper, we present an area efficient 
architecture of FFT processor that reuses the butterfly elements several 
times. The FFT processor is simulated using VHDL and the results are 
validated on a Virtex-6 FPGA. The proposed architecture outperforms the 
conventional architecture of a $N$-point FFT processor in terms of area 
which is reduced by a factor of $\log_N2$ with negligible increase in 
processing time.
 \end{abstract}

 \begin{IEEEkeywords}
 FFT, FPGA, Resource optimization
 \end{IEEEkeywords}

\IEEEpeerreviewmaketitle

 \section{Introduction}

 Field programmable gate arrays (FPGA) are programmed specifically for 
the problem to be solved, hence they can achieve higher performance with 
lower power consumption than general purpose processors. Therefore, FPGA 
is a promising implementation technology for computationally intensive 
applications such as signal, image, and network processing 
tasks~\cite{nordin}.

 Fast Fourier transform (FFT) is one of the most widely used operation 
in digital signal processing algorithms~\cite{IEEE:Baas} and plays a 
significant role in numerous signal processing applications, such as 
image processing, speech processing, software defined radio etc. FFT 
processors should be of higher throughput with lower computation time. 
So, for computing larger number of data samples, we have to think about 
the area of the FFT processor since the number of stage of FFT 
computation increases with a factor of $\log_2N$ . In the design of high 
throughput FFT architectures, energy-efficient design techniques can be 
used to maximize performance under power dissipation constraints.

Spatial and parallel FFT architecture, also known as array 
architecture~\cite{IEEE:you}, based on the Cooley-Tukey 
algorithm layout, is one of the potential high throughput designs. 
However, the implementation of the array architecture is hardware 
intensive. It achieves high performance by using spatial parallelism, 
while requiring more routing resources. However, as the problem size 
grows, unfolding the architecture spatially is not feasible due to 
serious power and area issue arisen by complex interconnections.

The pipelined architectures are useful for FFTs that require high data 
throughput~\cite{IEEE:he,IEEE:wold,IEEE:Torkelson,giri}. The basic 
principle of pipelined architectures is to collapse the rows. Radix-2 
multi-path delay commutator~\cite{IEEE:parhi}~\cite{lr} was probably the 
most classical approach for pipeline implementation of radix-2 FFT 
algorithm. Disadvantages include an increase in area due to the addition 
of memories and delay which is related to the memory 
usage~\cite{ouerhani}.

In this paper, we propose a novel architecture of area efficient FFT by 
reusing $N/2$ numbers of butterfly units more than once instead of using 
$(N/2)\log_2N$ butterfly units once~\cite{IEEE:chen}. This is achieved 
by a time control unit which sends back the previously computed data of 
$N/2$ butterfly units to itself for $(\log_2N)-1$ times and reuses the 
butterfly units to complete FFT computation. The area requirement is 
obviously smaller, only $N/2$ radix-$2$ elements, than the array 
architecture and pipelined architectures, $N$ being the number of sample 
points.

 \section{Traditional FFT Algorithm} 

 The Cooley-Tukey FFT algorithm is the most common algorithm for 
developing FFT. This algorithm uses a recursive way of solving FFT of 
any arbitrary size $N$. The technique divides the larger FFT into 
smaller FFTs which subsequently reduce the complexity of the algorithm. 
If the size of the FFT is $N$ then this algorithm makes $N=N1.N2$ where 
$N1$ and $N2$ are sizes of the smaller FFTs. Radix-2 decimation-in-time 
(DIT) is the most common form of the Cooley-Tukey algorithm, for any 
arbitrary size $N$. $N$ can be expressed as a power of 2, that is, $N = 
2^M$, where $M$ is an integer. This algorithm is called 
decimation-in-time since at each stage, the input sequence is divided 
into smaller sequences, i.e. the input sequences are decimated at each 
stage. A FFT of $N$-point discrete-time complex sequence $x(n)$, indexed 
by $n=0,1,....,N-1$ is defined as:
 \begin{equation} 
 Y(k)=\sum_{n=0}^{N-1}x(n)W_{N}^{nk}, k=0,1,...,N-1 
 \end{equation} 

\noindent where $W_N=e^{{-j2\pi}/N}$. Radix-2 divides the FFT into 
two equal parts. The first part calculates the Fourier transform of the 
even index numbers. The other part calculates the Fourier transform of 
the odd index numbers and then finally merges them to get the Fourier 
transform for the whole sequence.

Seperating the $x(n)$ into odd and even indexed values of $x(n)$, we 
obtain
 \begin{equation} 
 Y(k)={\sum_{n=0}^{{N/2}-1}x_e(n)W_{N/2}^{nk}}+W_N^k{\sum_{n=0}^{{N/2}-1}x_o(n)W_{N/2}^{nk}}
 \end{equation}

 \section{Proposed FFT Algorithm}

 The area of a FFT processor depends on the total number of butterfly 
units used. Each butterfly unit consists of multiplier and 
adder/subtractor blocks. Higher the bit resolution of samples, larger 
the area of these two mathematical blocks. According to traditional FFT 
algorithm each stage contains $N/2$ numbers of butterfly units. 
Therefore, for a traditional FFT processor, the total number of 
butterfly units is given by
 \begin{equation}
 BU_{Traditional FFT} = {(N/2)}{\log_2N}
 \end{equation}

 In the proposed algorithm, $N/2$ number of butterfly units are reused 
for $\log_2N$ times. Therefore, the modified architecture of FFT 
processor requires $BU_{Proposed FFT}$ number butterfly units which is 
given by 
 \begin{equation}
 BU_{Proposed FFT} = {N/2}
 \end{equation}. 

\noindent The proposed architecture of FFT processor reduces the number 
of butterfly units by a factor of ($\alpha$), which is given by
 \begin{equation}
 \begin{tabular}{rl}$\alpha$ & $= \frac{{N/2}}{{N/2}\log_2N}$ \\
& $={\log_2N}^{-1}$\\& $=\log_N2$ 
 \end{tabular}
 \end{equation} \ \ . 

Table~\ref{area} shows that the number of multipliers and 
adders/subtractors for the proposed FFT is less compared to that of 
the traditional FFT.
 \begin{table}[ht]
 \centering 
 \caption{Comparison of butterfly units, multipliers and adders/subtractors}
 \label{area}
 \begin{tabular}{|c|c|c|}\hline
& Traditional FFT & Proposed FFT \\ \hline
Butterfly unit (BU) & ${N/2}{\log_2N}$ & ${N/2}$ \\ \hline
Multiplier & ${N/2}\log_2N$ & $N/2$ \\ \hline
Adder/subtractor & $N\log_2N$ & $N$ \\ \hline
 \end{tabular}
 \end{table}
 \begin{table}[ht]
 \centering \caption{Number of butterfly units} \label{butterfly} 
 \begin{tabular}{|c|c|c|} \hline Number of samples & Traditional & Proposed\\
 & architecture & architecture\\ \hline
8 & 12 & 4\\ \hline
16 & 32 & 8\\ \hline
32 & 80 & 16\\ \hline
64 & 192 & 32\\ \hline
128 & 448 & 64\\ \hline
256 & 1024 & 128\\ \hline
512 & 2304 & 256\\ \hline
1024 & 5120 & 512\\ \hline
 \end{tabular}
 \end{table}
 \begin{figure}[ht]\centering
 \includegraphics[width=3.3in]{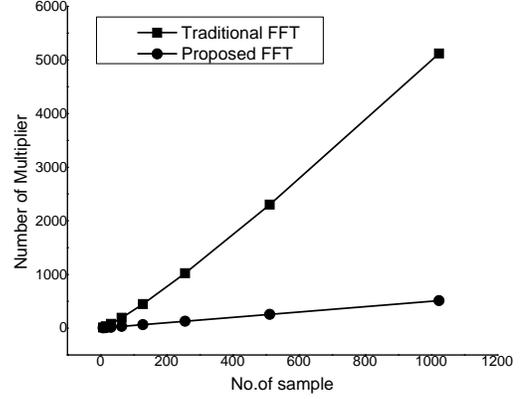}
 \caption{Comparison of number of multipliers required in traditional 
and proposed FFT processor}
 \label{numberofmuti}
 \end{figure}
 \begin{figure}[ht]\centering
 \includegraphics[width=3.3in]{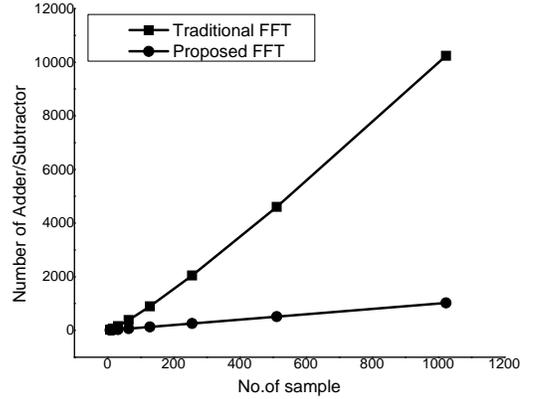}
 \caption{Comparison of number of adders/subtractors blocks required in 
traditional and proposed FFT processor}
 \label{numberofadd}
 \end{figure}

 \section{Architecture of Proposed FFT Processor}

\begin{figure}[ht]
 \centering
 \includegraphics[width=3.3in]{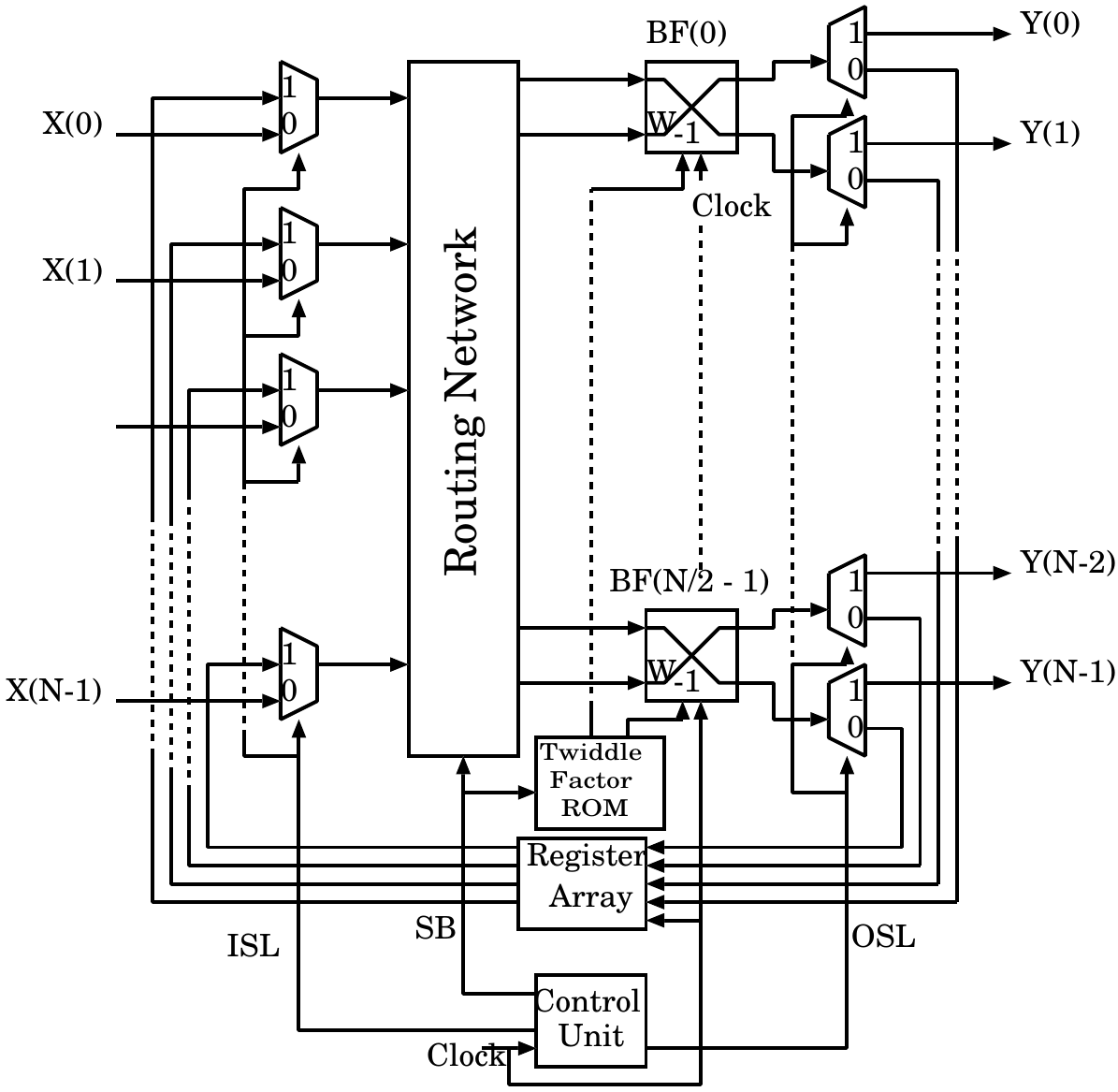}
 \caption{Architecture of the proposed FFT processor}
 \label{block of fft}
 \end{figure}

 The key feature of the proposed FFT processor is its low area. The 
proposed architecture reuses $N/2$ number of butterfly units for 
$\log_2N$ times. Block diagram of overall architecture of proposed FFT 
processor are shown in Fig.\ref{block of fft}. It consists of routing network, butterfly unit, control unit and input output enable 
blocks.

 \subsection{The control unit}
 Here the control unit is used for syncronizing all the blocks of the FFT 
processor. It counts the number of stage and controls the input, output and 
feedback databus. The control block generates three signals 1bit Input 
Select Line (ISL), output select line (OSL) and multi-bit stage bus 
(SB). This stage bus (SB) contains the stage of FFT computation. The control 
unit increments the number of stage with the rising edge of the clock 
signal. $ISL='0'$ at initial stage to select data from extarnal source 
after that $ISL='1'$ to select data from the feedback path or register 
array for $(\log_2N)-1$ times and $OSL='0'$ for $(\log_2N)-1$ times to 
fetch the output data of the butterfly unit to register array. At 
$\log_2N$th time $OSL='1'$ to enable the output data path of the FFT 
processor.

 \subsection{Butterfly unit (BU)}
\begin{figure}[ht]
 \centering
 \includegraphics[width=2in]{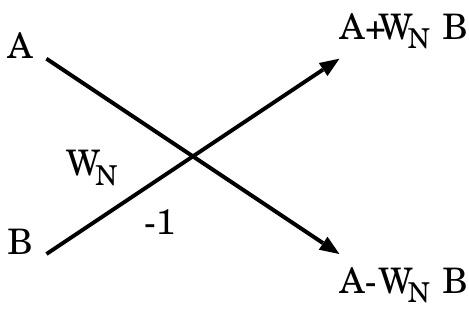}
 \caption{Architecture of butterfly unit}
 \label{2pointfft}
 \end{figure}

\begin{figure}[ht]
 \centering
 \includegraphics[width=2.5in]{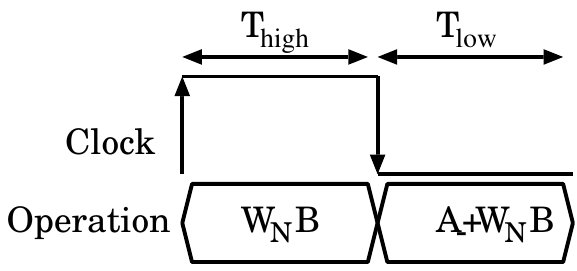}
 \caption{Timing diagram of butterfly unit}
 \label{timming2point}
 \end{figure}

 From the mathematical diagram, the output samples of the butterfly unit is 
generated after addition and subtraction operation with the product of 
even data sample and twiddle factor as shown in Fig.\ref{2pointfft}.
 These butterfly units are clock capable. The multiplication 
operation starts with the rising edge of the clock and addition or 
subtraction operation is done with the falling edge of the clock signal. 
So, that total operation of the butterfly unit is done within a single clock 
cycle as shown in Fig.\ref{timming2point}.

 \subsection{Twiddle factor ROM}
 The twiddle factor ROM stores the twiddle factor co-efficients. Size of 
this ROM unit is $\log_2N \times(N/2)$. This block have $N/2$ number of 
output signal which are connected with $N/2$ number of butterfly unit. 
The stage bus (SB) is connected with the address bus of ROM.

 \subsection{Routing network and register array}
\begin{figure}[ht]
 \centering
 \includegraphics[width=3.3in]{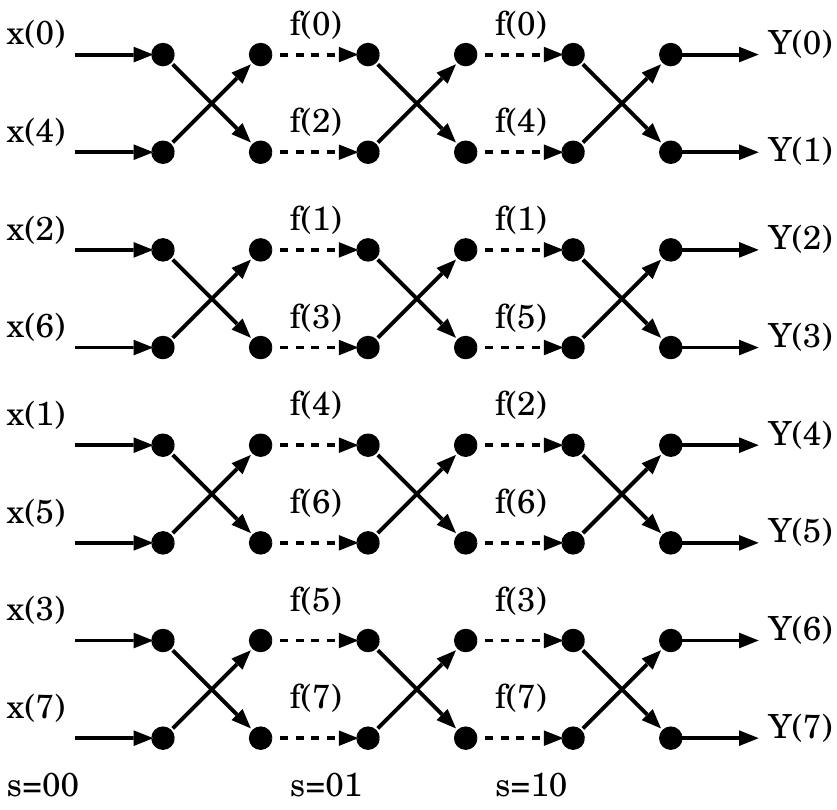}
 \caption{Data path layout of 8-point FFT processor}
 \label{route}
 \end{figure}

 Routing network unit passes the proper sequences of input samples to 
the butterfly units for different stage of computations. Value of stage 
bus (SB) controls output of this unit.  At first stage, routing network 
generates bit-reversed sample sequence of input samples. For the remain 
stages, the routing network shuffles the feedback samples with distance of 
$2^{m-1}$ where $m=2,3,...{\log_2N}$. Figure.\ref{route} shows the data 
path layout of 8-point FFT. Dashed arrows define the feedback samples. Cross arrows signifies the butterfly units. Register array~\cite{IEEE:wu} holds the previous data of the butterfly 
units and passes the stored data with the rising edge of the clock.

 \section{Implementation and Results}

\begin{figure*}\centering
 \includegraphics[width=\textwidth]{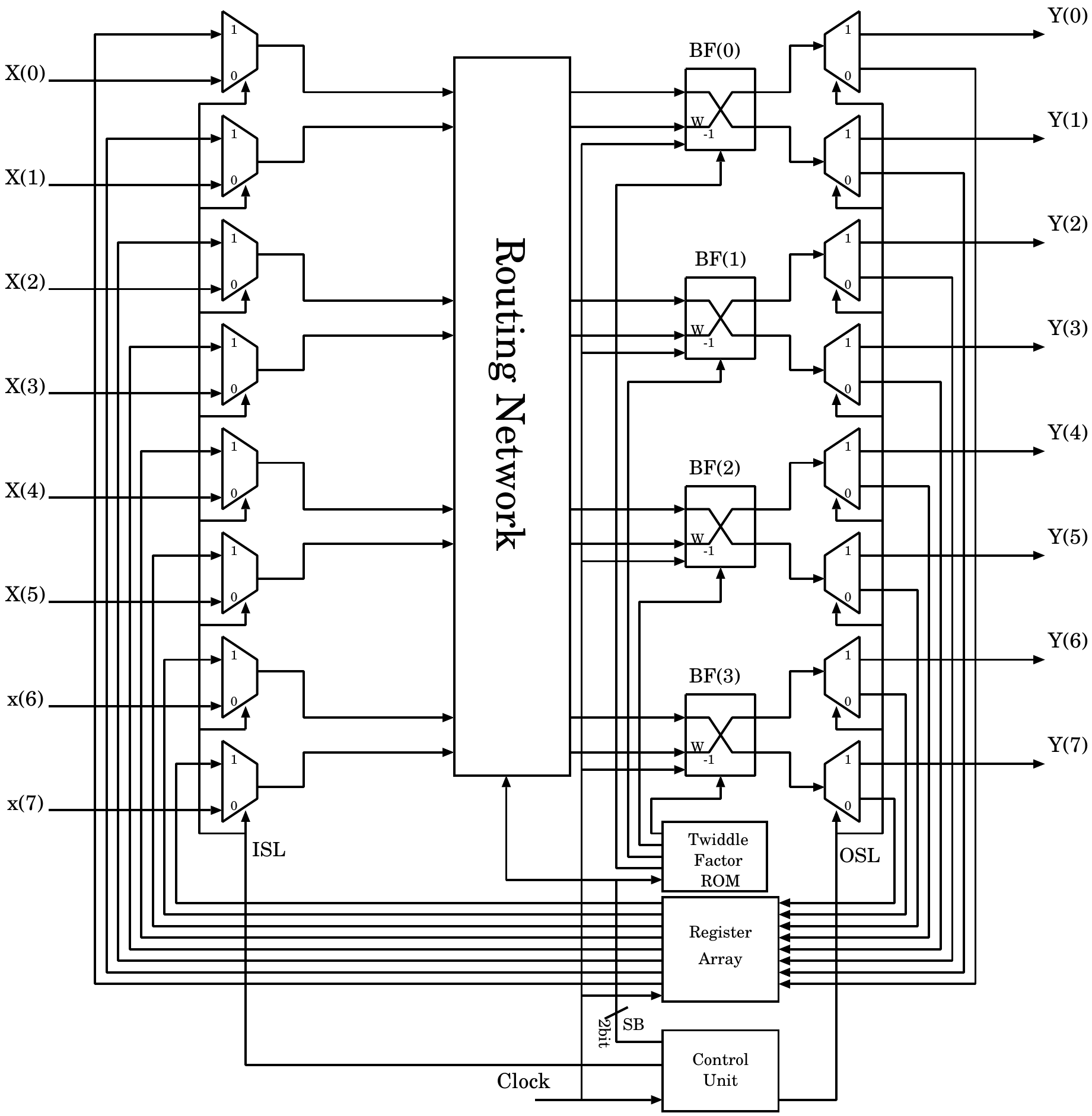}
 \caption{Proposed architecture of 8-point FFT processor}
 \label{fft}
 \end{figure*}

\begin{figure}[ht]
 \centering
 \includegraphics[width=3in]{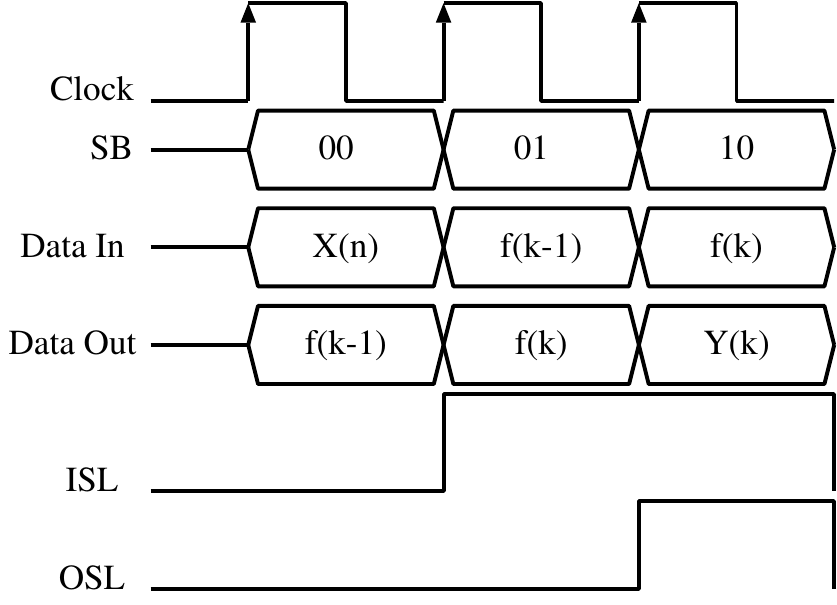}
 \caption{Timming Diagram of 8-point FFT}
 \label{timming}
 \end{figure}

 Figure \ref{fft} shows the architecture of 8-point FFT 
processor according to proposed FFT processor. Figure.\ref{timming} shows 
the timming diagram of this processor. $X(n)$ and $Y(K)$ denotes input 
and output samples and $f(k)$ are the ouput samples of previous stage.

\begin{figure}[ht]
 \centering
 \includegraphics[width=3.6in]{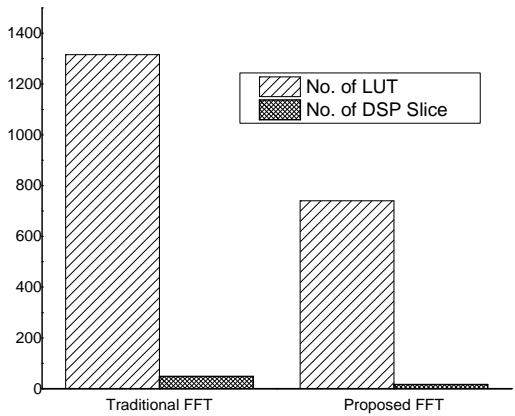}
 \caption{Comparison of number LUT and DSP slice between traditional 
and proposed FFT processor}
 \label{Re}
 \end{figure}

\begin{figure}[ht]
 \centering
 \includegraphics[width=3.6in, trim={0 8 0 8.5},clip]{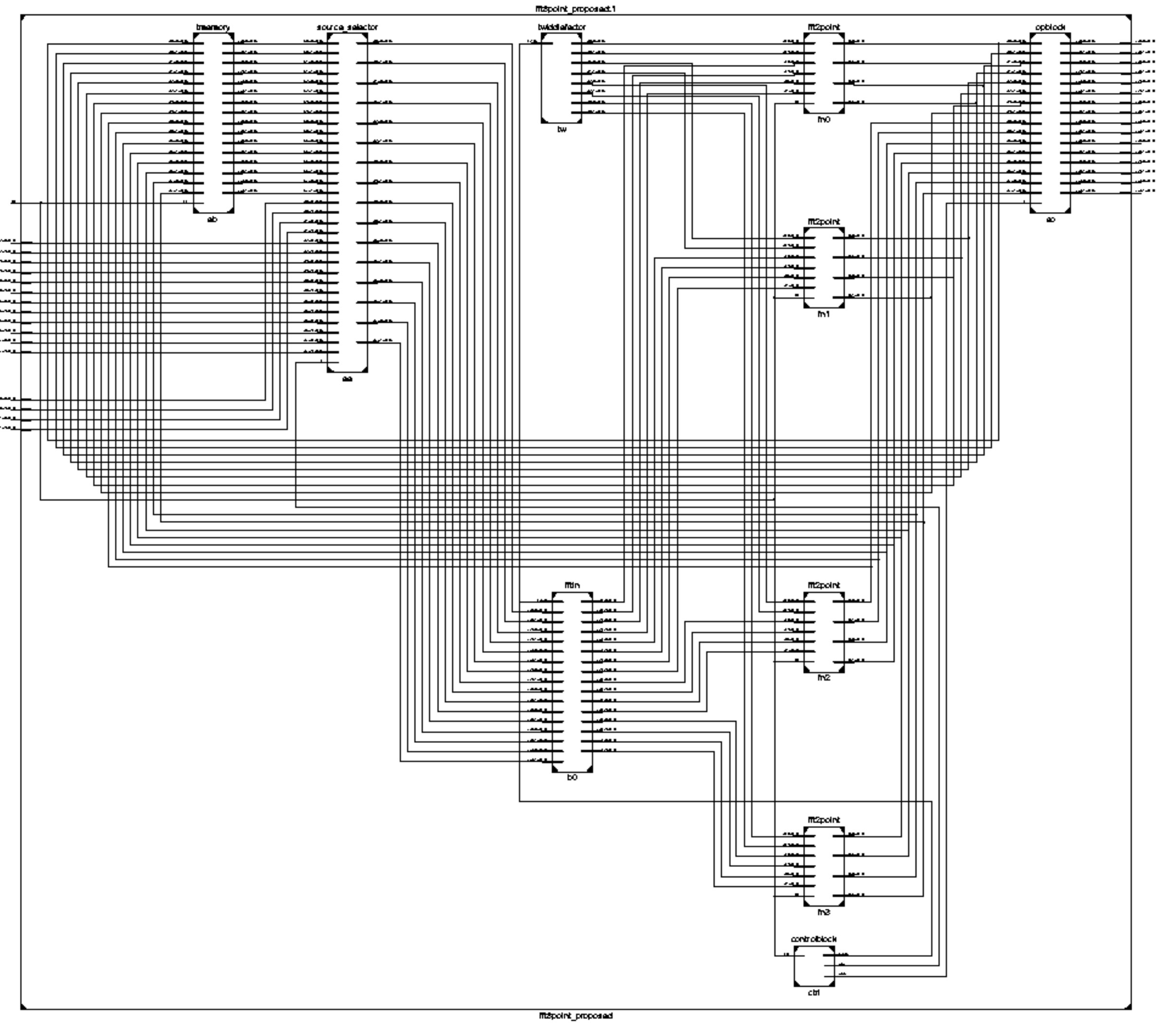}
 \caption{Details of RTL diagram of 8-point FFT processor}
 \label{RTL}
 \end{figure}

 The proposed architecture for 8-point FFT processor is coded using 
VHDL, emulated and synthesized using Xilinx ISE 14.2 for Virtex-6 FPGA. 
Table.\ref{ha} shows the comparison of advanced HDL synthesis reports 
with traditional FFT. Figure.\ref{RTL} shows the generated detailed RTL 
diagram of proposed 8-point FFT processor. Figure.\ref{Re} shows the 
comparison of number of DSP slices and LUTs requirements and 
Table.\ref{de} shows the comparison of timming delay with the 
traditional FFT processor.
 \begin{table}[ht]
 \centering
 \caption{Comparison of advanced HDL synthesis reports}
 \label{ha}
 \begin{tabular}{|c|c|c|}\hline
Hardware & Traditional FFT & Proposed FFT \\ \hline
MACs & 24 & 8 \\ \hline
Multipliers & 24 & 8 \\ \hline
Adder/Subtractors & 72 & 25 \\ \hline
Multiplexers & 360 & 136 \\ \hline
XORs & 24 & 8 \\ \hline
Registers & -- & 288 \\ \hline
counter & -- & 1 \\ \hline
 \end{tabular}
 \end{table}
 \begin{table}[H]
 \centering
 \caption{Comparison of delay between traditional and proposed FFT processor}
 \label{de}
 \begin{tabular}{|c|c|} \hline
 Algorithm & Delay (nsec)\\ \hline
 Traditional FFT & 29.111\\ \hline
 Proposed FFT & 29.397\\ \hline
 \end{tabular}
 \end{table}

 \begin{table}[H]
 \centering
 \caption{Device utilization and timing summary}
 \label{summary}
 \begin{tabular}{|c|c|} \hline
 \multicolumn{2}{|c|}{Device Utilization Summary}\\ \hline
 Selected Device & 6vsx475tff1759-2\\ \hline
 Number of Slice Registers & 301  out of  595200\\ \hline
 Number of Slice LUTs & 748  out of  297600\\ \hline
 Number of DSP48E1s & 16  out of 2016\\ \hline
 \multicolumn{2}{|c|}{Timing Summary}\\ \hline
 Minimum period & 19.598ns \\ \hline
 Maximum Frequency & 51.025MHz \\ \hline
 Minimum input arrival & 9.384ns\\ 
 time before clock & \\ \hline
 Maximum output required & 0.665ns\\ 
 time after clock & \\ \hline
 \end{tabular}
 \end{table}

 \section{Conclusion}

 The proposed architecture presents an area efficient Radix-2 
FFT processor. The algorithm reuses the butterfly units of single stage 
more than once which reduces the area drastically. The architecture has 
been emulated and the performance analysis has been carried out in 
terms of overall response time and utilization of hardware resources of 
FPGA. Detailed analysis reveals that the proposed architecture reduces 
the area dramatically without compromising the response time. Further 
improvements may be obtained by designing silicon layout and analysing 
the post-layout performance trade-off.

 \balance

\end{document}